\documentclass{article}

\usepackage{amsmath}
\makeatletter 
\@addtoreset{equation}{section}
\makeatother  

\newtheorem{theorem}{\hspace{2em}Theorem}
\newtheorem{lemma}{\hspace{2em}Lemma}
\usepackage{indentfirst}
\usepackage{setspace}
\usepackage{amsmath}
\usepackage{cases}
\usepackage[colorlinks, linkcolor=black, anchorcolor=black, citecolor=black]{hyperref}
\usepackage{graphicx}
\usepackage{tocloft}
\title{\textbf{Angular Momentum-Free of the Entropy Relations for Rotating Kaluza-Klein Black Holes}}
\author{Hang Liu$^{a}$\thanks{E-mails:hangliu@mail.nankai.edu.cn} and Xin-he Meng$^{a,b}$\thanks{E-mails:xhm@nankai.edu.cn}
\\
\\
$^{a}$ School of Physics, Nankai University, Tianjin 300071, China\\
$^{b}$ State Key Laboratory of Theoretical Physics, Institute of Theoretical Physics,\\
Chinese Academy of Science, Beijing 100190, China}
\date{}
\begin{document}
\large
\maketitle
\begin{abstract}
Based on a mathematical lemma related to the Vandermonde determinant and two theorems derived from the first law of black hole thermodynamics, we investigate the angular momentum independence of the entropy sum as well as the entropy product of general rotating Kaluza-Klein black holes in higher dimensions. We show that for both non-charged rotating Kaluza-Klein black holes and non-charged rotating Kaluza-Klein-AdS black holes, the angular momentum of the black holes will not be present in entropy sum relation in dimensions $d\geq4$, while the independence of angular momentum of the entropy product holds provided that the black holes possess at least one zero rotation parameter $a_j$ = 0 in higher dimensions $d\geq5$, which means that the cosmological constant does not affect the angular momentum-free property of entropy sum and entropy product under the circumstances that charge $\delta=0$. For the reason that the entropy relations of charged rotating Kaluza-Klein black holes as well as the non-charged rotating Kaluza-Klein black holes in asymptotically flat spacetime act the same way, it is found that the charge has no effect in the angular momentum-independence of entropy sum and product in asymptotically flat spactime.
\end{abstract}
\tableofcontents
\section{Introduction}
Since Hawking and Bekenstein \textit{et al.} had established black hole thermodynamics in their pioneer work by Hawking's radiation with black body spectrum and in analogy between the black hole mechanics with the classical thermodynamics, studying black hole entropy has attracted much attention and it is still a major challenge to understand the origin of the black hole entropy at microscopic level in quantum gravity (though we have not achieved the final theory over years' efforts). In recent years, more interests have been focused on the study of the mass-independence of entropy sum and entropy product for stationary black holes with more than one horizons and people find that the product of all horizon entropies is mass independent in many cases \cite{Cvetic1,Vandoren1,Lu1,Shoom1,Lu2,Chow1, Rodriguez1,Visser1,Chenbin1}, but fails  in some cases \cite{Detournay1,Visser2,Moreno1, Kastor1}. However, it is found by Wang, Xu and Meng \cite{Meng1} that the entropy sum of all horizons, including ``virtual horizons", is a more  universal relation which is also mass independent and only depends on the coupling constants of the theory and topology of black holes in asymptotical Kerr-Newman-(anti)-dS spacetime background, in most cases, which means that it is more general than entropy product in some cases. While in  asymptotical flat spacetime background, entropy product may be more general than entropy sum, e.g., for Myers-Perry black holes, the mass-independence of entropy product holds in all dimensions while the  entropy sum is independent of mass only in higher dimensions $d\geq 5$, which has been demonstrated in ref.\cite{Gao1}. But if we take the angular momentum of a rotating black hole into consideration, we find that the entropy sum is  independent of angular momentum in all dimensions while the entropy product is angular mentum free if the black hole possesses at least one rotation parameter $a_j=0$ in higher dimensions $d\geq 5$, as  ref.\cite{Hang} has showed. In this sense, the entropy sum relation is still more general than entropy product relation.

With the discovery of the remarkable anti-de Sitter/conformal field theory(AdS/CFT) correspondence, people have made a great progress in understanding of string theory and quantum gravity over the last few years, and the new significance of the investigation of the solution of gauged supergravities has been acquired. It is of considerable interest to study the rotating charged black holes with a cosmological constant in higher dimensions since such black holes can provide new important gravitational backgrounds for understanding the micro entropy of black holes and testing the AdS/CFT correspondence within the string theory framework. The solution in Kaluza-Klein supergravity of general rotating charged Kaluza-Klein-AdS black holes in higher dimensions has been obtained by author in Ref.\cite{Wu1} who has showed that the conserved charges of this black holes satisfy the first law of thermodynamics and also explained that the general Kaluza-Klein-AdS solutions have a beautiful structure similar to the Kerr-Schild ansatz, which highlights its promising application to include multiple electric charges into solutions yet to be discovered in gauged supergravity.

To have a deeper understanding of black holes, it is important to study the entropy product of all horizons \cite{Cvetic1}, especially for the black holes which only have two horizons,i.e., outer event horizon and inner Cauchy horizon, and \cite{Chenbin1,Chenbin2} show that the inner horizon plays an important role in the study of black hole physics. For general four dimensions(4D) and five dimensions(5D) multi-charged rotating black holes, the entropy of the outer and inner horizons can be expressed in terms of $N_{L}$ and $N_{R}$, which are interpreted as the level of left-moving and right-moving excitations of a two dimensional CFT \cite{ Larsen1, Larsen2,Cve}, and the entropies are
\begin{equation}
S_{\pm}=2\pi(\sqrt{N_L}\pm\sqrt{N_{R}})
\end{equation}
where we denote the entropy of outer horizon and inner horizon as  $S_{+}$ and $S_{-}$, respectively. For the entropy product, we have
\begin{equation}
S_{+}S_{-}=4\pi^2(N_{L}^2-N_{R}^2)
\end{equation}
which should be quantized and must be mass independent, and  being only expressed as the function of quantized angular momentum and other charges. Actually, For an axisymmetric and stationary Einstein-Maxwell black hole in four dimensions with angular momentum J and charge Q, Marcus Ansorg and Jorg Hennig \cite{Ansorg,Hennig} have proved the universal relation below
\begin{equation}
A_+A_- = (8\pi J)^2+(4\pi Q^2)^2
\end{equation}
where $A_+$ and $A_-$ represent the area of event horizon(outer horizon)and Cauchy horizon(inner horizon), respectively.
Cveti\ifmmode \check{c}\else \v{c}\fi{} \textit{et al.} \cite{Cvetic1} have generalized the investigations to rotating black holes with multi-horizons in higher dimensions by explicit calculations. Authors in ref.\cite{Cvetic1} have demonstrated that the entropy product of all horizons for rotating multi-charge black holes in four and higher dimensions is mass free and only depends on its charges $Q_i$ and angular momentum $J_j$ in either asymptotically flat or asymptotically AdS spacetime, while the ref.\cite{Hang} has proved that the entropy product can also be angular momentum free if the rotating black hole owns at least one rotation parameter $a_j=0$ in higher dimensions $d\geq5$.

So far, as we have introduced above, the mass-independence of the entropy sum and entropy product of all horizons in variety of theories of gravity, including modified gravity theories \cite{Meng3,Hang2}, have drawn much attention while the angular momentum-independence of the entropy sum and product is merely investigated. In this paper,  we keep on studying the mathematical physics related to mass and angular momentum-free relations for black holes, which is motivated as the following consideration.
We know that for a massive star with mass $M$ large enough, angular momentum $J$ and charge $Q$ will collapse or two massive black holes merge into a black hole, i.e., Kerr-Newman black hole after burning out all its nuclear fuel, and then all the parameters needed to describe it are just its mass, angular momentum and charge mathematically. This fact is summarized  as the famous ``no hair or three hairs theorem for black hole". In astrophysics, physical black holes only possess only two independent parameters which are its mass and angular momentum as its final state characters. If there exist some relations which are independent of angular momentum (and/or mass-free for more complete case) of a rotating star/black hole, we say these relations are universal, which means no matter what initial conditions it starts with the final state is universal (mass or/and angular momentum free). Motivated by this thought, we keep on studying the angular momentum-independence, and we generalize our discussions  to rotating Kaluza-Klein black holes, including the charged case, which is not involved in our previous work  \cite{Hang}. In this work, by applying  a mathematical lemma related to the Vandermonde determinant and the first law of black hole thermodynamics, we find that for both non-charged rotating Kaluza-Klein black holes and non-charged Kaluza-Klein-AdS black holes, the entropy sum is always angular momentum independent in dimensions $d\geq4$, and the  entropy product is angular momentum free if the black holes possess at least one zero rotation parameter $a_j$ = 0 in higher dimensions $d\geq5$, which means that the cosmological constant does not affect the angular momentum-independence property of entropy sum and entropy product if  the black holes possess no charge. For charged rotating Kaluza-Klein black holes in asymptotically flat spactime, we get the same conclusions as that in non-charged case, which means that the charge has no effect in the angular momentum-independence of entropy sum and product in asymptotically flat spactime.

This paper is organized as follows. In next section, we will establish a criterion for the angular momentum-independence of the entropy sum and product by employing the first law and a mathematical lemma. In section 3.1, we will introduce the general rotating Kaluza-Klein black holes and its thermodynamical quantities. In section 3.2 and 3.3, we will study the angular momentum-independence of the entropy sum and entropy product by using the mathematical lemma and the two theorems given in section 2, respectively. The last section is devoted to conclusions and discussions.

\section{Mathematical Lemma and the First Law of Thermodynamics}
In this section, we introduce a mathematical lemma related to the Vandermonde determinant and two theorems based on the first law of black hole thermodynamics which are all critical to our arguments as we show in the following sections. The proof of this lemma can be found in  ref.\cite{Gao1}. We will index  which horizon by subscript $i$ and index which angular momentum by subscript $j$ throughout the paper.
\subsection{The First Law of Thermodynamics}
 For a stationary rotating multi-horizons black hole with mass $M$ , electric charge $Q$ and angular momentum $J_j$ in the spacetime with a cosmological constant $\Lambda$, each horizon possesses corresponding Hawking temperature $T_i$, entropy $S_i$, angular velocity $\Omega_{ij}$, electric potential $\Phi_i$ and thermodynamics volume $V_i$ related to cosmological constant which we treat it as a dynamical variable \cite{D1,B1,M1,B2,N1,Meng4}. The first law of thermodynamics for each horizon reads
\begin{equation}
dM=T_idS_i+\Phi_i dQ+\sum_{j=1}^{m}\Omega_{ij}dJ_j+V_id\Lambda
\end{equation}
which yields
\begin{equation}
dS_i=\frac{1}{T_i}(dM-\Phi_i dQ-\sum_{j=1}^{m}\Omega_{ij}dJ_j-V_id\Lambda)\label{1}
\end{equation}
Obviously, the E.q (\ref{1}) leads to following relation
\begin{equation}
\frac{\partial S_i}{\partial J_j}=-\frac{\Omega_{ij}}{T_i}
\end{equation}
For the entropy sum $\widetilde{S}=S_1+S_2+S_3+\ldots +S_n$, we have
\begin{equation}
\frac{\partial\widetilde{S}}{\partial J_j}=-\sum_{i=1}^{n}\frac{\Omega_{ij}}{T_i}
\end{equation}
For the entropy product $\widehat{S}=S_1S_2S_3\ldots S_n$, it's not hard to get
\begin{equation}
\frac{\partial\widehat{S}}{\partial J_j}=-\widehat{S}\sum_{i=1}^{n}\frac{\Omega_{ij}}{S_i T_i}
\end{equation}
we require $T_i\neq0$ and $S_i\neq0$ in this paper.

From the calculations above, we arrive at two theorems
\begin{theorem}\label{theorem1}
For an axisymmetric and stationary rotating black hole with $n\geq2$ horizons and possesses $m\geq1$ rotation parameters $a_j$, the entropy sum of all horizons is independent of angular momentum if and only if
\begin{equation}
\frac{\partial{\widetilde S}}{\partial J_j}=-\sum_{i=1}^{n}\frac{\Omega_{ij}}{T_i}=0,(j=1,2,3,\ldots,m)
\end{equation}
\end{theorem}
\begin{theorem}\label{theorem2}
For an axisymmetric and stationary rotating black hole with $n\geq2$ horizons and possesses $m\geq1$ rotation parameters $a_j$, the entropy product of all horizons is independent of angular momentum if and only if
\begin{equation}
\sum_{i=1}^{n}\frac{\Omega_{ij}}{T_i S_i}=0,(j=1,2,3,\ldots ,m)
\end{equation}
\end{theorem}

\subsection{Mathematical Lemma}
\begin{lemma}\label{lemma}
Let \{$r_i$\} be n different numbers, then
\begin{equation}
\sum_{i=1}^{n}\frac{r_i^k}{\prod\limits_{j\neq i}^{n}(r_i-r_j)}=0
\end{equation}
where $0\leq k\leq n-2$
\end{lemma}

The proof of the lemma is a simple application of the Vandermonde determinant and the general proof is given in ref.\cite{Gao1}. One can test the validity of this lemma. For example, when $n=3$, it gives
\begin{gather}
\frac{1}{(r_1-r_2)(r_1-r_3)}+\frac{1}{(r_2-r_1)(r_2-r_3)}+\frac{1}{(r_3-r_1)(r_3-r_2)}=0\\
\frac{r_1}{(r_1-r_2)(r_1-r_3)}+\frac{r_2}{(r_2-r_1)(r_2-r_3)}+\frac{r_3}{(r_3-r_1)(r_3-r_2)}=0
\end{gather}
while
\begin{gather}
\frac{r_1^2}{(r_1-r_2)(r_1-r_3)}+\frac{r_2^2}{(r_2-r_1)(r_2-r_3)}+\frac{r_3^2}{(r_3-r_1)(r_3-r_2)}=1\\
\frac{r_1^3}{(r_1-r_2)(r_1-r_3)}+\frac{r_2^3}{(r_2-r_1)(r_2-r_3)}+\frac{r_3^2}{(r_3-r_1)(r_3-r_2)}=r_1+r_2+r_3
\end{gather}

\section{Entropy Sum and Entropy Product of the Rotating Kaluza-Klein Black Holes}
In this section, we will investigate the angular momentum independence of the entropy sum and entropy product of all horizons of general rotating Kaluza-Klein black holes in higher dimensions($d\geq4$) based on the theorems and lemma we introduce in section two.
\subsection{General Rotating Charged Kaluza-Klein-AdS Black Holes}
Suppose the spacetime dimensions $d=2n+1+\epsilon\geq4$, with $n=[(d-1)/2]$ being the number of rotating parameters $a_j$ and $2\epsilon=1+(-1)^d$. The metric form in Boyer-Lindquist coordinates of general rotating charged Kaluza-Klein-AdS black holes can be  written as \cite{Wu1,Pop2}
\begin{equation}
\begin{split}
ds^2=H^{1/(d-2)}&\left[-(1+g^2r^2)Wdt^2+\frac{Udr^2}{V(r)-2mf(r)}
+\sum_{i=1}^{n+\epsilon}\frac{r^2+a_i^2}{\chi_i}d\mu_i^2\right.\\
&\left.\quad+\sum_{i=1}^{n}\frac{r^2+a_i^2}{\chi_i}\mu_i^2 d\phi_i^2
-\frac{g^2}{(1+g^2r^2)W}\left(\sum_{i=1}^{n+\epsilon}\frac{r^2+a_i^2}{\chi_i}\mu_id\mu_i \right)^2\right.\\
&\left.\quad+\frac{2m}{UH}\left( cWdt-\sum_{i=1}^{n}\frac{a_i\sqrt{}\Xi_i}{\chi_i}\mu_i^2d\phi_i\right)^2\right]\label{metric}
\end{split}
\end{equation}
where $\mu_i$ is restricted as $\sum_{i=1}^{n+\epsilon}\mu_i^2=1$ and $0\leq\mu_i\leq 1$
for $1\leq i\leq n$, while(for even d) $-1\leq \mu_{n+1}\leq 1$ and $a_{n+1}=0$. The functions $(U,W,F,H,V(r))$ and $f(r)$ are defined to be
\begin{gather}
U=r^{\epsilon}\sum_{i=1}^{n+\epsilon}\frac{\mu_i^2}{r^2+a_i^2}\prod\limits_{j=1}^{n}(r^2+a_j^2),
\quad W=\sum_{i=1}^{n+\epsilon}\frac{\mu_i^2}{\chi_i}\\
F=\frac{r^2}{1+g^2r^2}\sum_{i=1}^{n+\epsilon}\frac{\mu_i^2}{r^2+a_i^2},
\qquad H=1+\frac{2ms^2}{U}\\
f(r)=c^2-s^2(1+g^2r^2),\qquad \Xi_i=c^2-s^2\chi_i
\end{gather}
and
\begin{equation}
\chi_i=1-g^2a_i^2,\quad c=\cosh{\delta},\quad s=\sinh{\delta}
\end{equation}
The parameter $\delta$ represents charge and parameter $g$ is defined as
\begin{equation}
g^2=-\frac{2\Lambda}{(d-1)(d-2)}
\end{equation}

$\Lambda$ is cosmological constant. The metric above reduces to metric derived in ref.\cite{Pope1,Pop2} and the case in ref.\cite{Kunz1} when charge $\delta=0$ and cosmological constant $\Lambda=0$, respectively. The author in ref.\cite{Wu1} has checked that the general solution (\ref{metric}) satisfies the field equations derived from lagrangian of the Einstein-Maxwell-Dilaton system for the $d=4,5,6,7$ cases.

The surface gravity on the i-th horizon of the rotating KK-AdS black holes is given by \cite{Wu1}
\begin{equation}
\begin{split}
k_i&=\frac{r_i(1+g^2r_i^2)\sqrt{f(r_i)}}{c}\left[\sum_{j=1}^{n}\frac{1}{r_i^2+a_j^2}
+\frac{\epsilon-2}{2r_i^2}+\frac{g^2c^2}{(1+g^2r_i^2)f(r_i)}
\right]\\
&=\frac{(1+g^2r_i^2)^2}{4m\sqrt{f(r_i)}cr_i^{2-\epsilon}}\frac{\partial h(r)}{\partial r}\bigg |_{r=r_i}
\end{split}
\end{equation}
where $r_i$ stands for the radius of corresponding horizon and the function $h(r)$ is the horizon function which is
\begin{gather}
h(r)=\prod\limits_{j=1}^{n}(r^2+a_j^2)-\frac{2mf(r)r^{2-\epsilon}}{1+g^2r^2}\label{h(r)}\\
h(r_i)=0
\end{gather}
the $m$ is mass parameter. Hence the Hawking temperature on the i-th horizon is
\begin{equation}
T_i=\frac{k_i}{2\pi}=\frac{(1+g^2r_i^2)^2}{8\pi m\sqrt{f(r_i)}cr_i^{2-\epsilon}}\partial_r{\left(\prod\limits_{j=1}^{n}(r^2+a_j^2)-\frac{2mf(r)r^{2-\epsilon}}{1+g^2r^2}\right)}\bigg |_{r=r_i}\label{2}
\end{equation}

The entropy of the i-th horizon is evaluated as
\begin{equation}
S_i=\frac{V_{d-2}r_i^{\epsilon-1}c}{4\sqrt{f(r_i)}}\prod\limits_{j=1}^{n}\frac{r_i^2+a_j^2}{\chi_j}\label{entropy}
\end{equation}
where $\chi_j=1-g^2a_j^2\neq0$ and the volume of unit $(d-2)$-sphere is denoted as
\begin{equation}
V_{d-2}=\frac{2\pi^{(d-1)/2}}{\Gamma[(d-1)/2]}
\end{equation}

The angular velocity corresponding to the i-th horizon and j-th rotation parameter $a_j$ is given by
\begin{equation}
\Omega_{ij}=\frac{(1+g^2r_i^2)a_j\sqrt{\Xi_j}}{(r_i^2+a_j^2)c}\label{angular}
\end{equation}
and the rotation parameters $a_j$ are related to the angular momentum $J_j$, which appear in the first law, via
\begin{equation}
J_j=\frac{V_{d-2}ma_j c\sqrt{\Xi_j}}{4\pi \chi_j \prod_{i=1}^{n}\chi_i}
\end{equation}
where $\Xi_j=c^2-s^2\chi_j=c^2-s^2(1-g^2a_j^2)=1+s^2g^2a_j^2\neq0$.

\subsection{Entropy Sum}
We adopt the similar procedure which was used in ref.\cite{Gao1} to discuss the angular momentum independence of the entropy sum.

We introduce a function
\begin{equation}
F(r)=\frac{(1+g^2r^2)^2}{4m\sqrt{f(r)}cr^{2-\epsilon}}h(r)=\frac{(1+g^2r^2)^2}{4m\sqrt{f(r)}cr^{2-\epsilon}}
\prod\limits_{j=1}^{2n+\lambda}(r-r_j)\label{F(r)}
\end{equation}
the last term of E.q (\ref{F(r)}) holds since the horizon function $h(r)$ has $(2n+\lambda)$ roots, and
\begin{numcases}{\lambda=}
 2 & $ g^2>0$ \\
 0 & $ g^2=0$
 \end{numcases}
Then we take derivative of $F(r)$ with respect to $r$, we get
\begin{equation}
F'(r_i)=\frac{(1+g^2r_i^2)^2}{4m\sqrt{f(r_i)}cr_i^{2-\epsilon}}\partial_{r}h(r)|_{r=r_i}=\frac{(1+g^2r_i^2)^2}{4m\sqrt{f(r_i)}cr_i^{2-\epsilon}}
\prod\limits_{j\neq i}^{2n+\lambda}(r_i-r_j)\label{3}
\end{equation}
where $r_i$ represents the horizon radius. Under the consideration of E.q (\ref{2}), we have
\begin{equation}
F'(r_i)=2\pi T_i\label{4}
\end{equation}
together with E.q (\ref{3}) and (\ref{4}), we obtain
\begin{equation}
\frac{1}{T_i}=\frac{8\pi m \sqrt{f(r_i)}cr_i^{2-\epsilon}}{(1+g^2r_i^2)^2\prod\limits_{j\neq i}^{2n+\lambda}(r_i-r_j)}\label{16}
\end{equation}
By using angular velocity formula (\ref{angular}), we get
\begin{equation}
\sum_{i=1}^{2n+\lambda}\frac{\Omega_{ij}}{T_i}=8\pi ma_j\sqrt{\Xi_j}\sum_{i=1}^{2n+\lambda}\frac{\sqrt{f(r_i)}r_i^{2-\epsilon}}{(1+g^2r_i^2)\prod\limits_{j\neq i}^{n}(r_i-r_j)}\frac{1}{r_i^2+a_j^2}\label{6}
\end{equation}
Note that we have horizon function (\ref{h(r)}), which yields
\begin{equation}
\frac{1}{r_i^2+a_j^2}=\frac{(1+g^2r_i^2)}{2mf(r_i)r_i^{2-\epsilon}}\prod\limits_{k\neq j}^{n}(r_i^2+a_k^2)\label{5}
\end{equation}
substituting E.q (\ref{5}) into E.q (\ref{6}), we finally obtain
\begin{equation}
\sum_{i=1}^{2n+\lambda}\frac{\Omega_{ij}}{T_i}=4\pi a_j\sqrt{\Xi_j}\sum_{i=1}^{2n+\lambda}\frac{\prod\limits_{k\neq j}^{n}(r_i^2+a_k^2)}{\sqrt{f(r_i)}\prod\limits_{j\neq i}^{2n+\lambda}(r_i-r_j)}\label{7}
\end{equation}

In odd dimensions, the horizon function (\ref{h(r)}) is a function of $r^2$ and the entropy function $S_i$ is a function of $r_i$ with odd power so that the each root $r_i$ of function $h(r)$  is coupled with $-r_i(i=1,2,3,\ldots,n+\lambda/2)$ which make the entropy sum $\widetilde S$ vanish, just as \cite{Meng1,Meng2,Meng5} have showed. However, when it comes to E.q (\ref{7}), the numerator $\prod_{k\neq j}^{n}(r_i^2+a_k^2)$ only consists of $r_i$ with even power and the denominator $\sqrt{f(r_i)}\prod_{j\neq i}^{2n+\lambda}(r_i-r_j)=-\sqrt{f(r_i)}\prod_{j\neq i+1}^{2n+\lambda}(r_{i+1}-r_j)$ with $i=1,3,5,\ldots,2n+1$ and we let $r_i$ and $r_{i+1}$ are coupled roots which meet the condition $r_i+r_{i+1}=0$ in odd dimensions. Note that we have $f(r_i)=f(-r_i)$. Under this consideration, the E.q (\ref{7}) has a value of zero. For example, we set $r_1=-r_2$, $r_3=-r_4$, then
\begin{gather}
\frac{r_1^2}{\sqrt{f(r_1)}(r_1-r_2)(r_1-r_3)(r_1-r_4))}+\frac{r_2^2}{\sqrt{f(r_2)}(r_2-r_1)(r_2-r_3)(r_2-r_4)}=0\\
\frac{r_3^2}{\sqrt{f(r_3)}(r_3-r_1)(r_3-r_2)(r_3-r_4)}+\frac{r_4^2}{\sqrt{f(r_4)}(r_4-r_1)(r_4-r_2)(r_4-r_3)}=0
\intertext{the equations above can be written as a general and unified form}
\sum_{i=1}^{2n}\frac{\xi(r_i)r_i^{2\kappa}}{\prod_{j\neq i}^{j=2n}(r_i-r_j)}=0,\{\xi(r_i)=\xi(-r_i), r_1=-r_2,r_3=-r_4,\ldots\}\label{15}
\end{gather}
where $\kappa$ is an integer. By employing theorem (\ref{theorem1}) and the note that the zero value of E.q (\ref{7}), we conclude that the entropy sum of all horizons of general rotating KK-AdS black holes is angular momentum independent in odd dimensions. This result just coincides  with  the fact that the entropy sum $ \widetilde S=0$ of all horizons in odd dimensions.

In even dimensions, it's another matter and we must use that lemma (\ref{lemma}) we introduce in section two to investigate the value of E.q (\ref{7}) is zero or non-zero. We would like to  carry out Taylor series expansion for function $f(r_i)^{-1/2}$ as
\begin{equation}
\frac{1}{\sqrt{f(r_i)}}=1+\frac{1}{2}g^2s^2r_i^2+\frac{3}{8}g^2s^2r_i^4+
\frac{5}{16}g^6s^6r_i^6+\dots\label{8}
\end{equation}
After substituting (\ref{8}) into (\ref{7}), on can easily find that E.q (\ref{7}) vanishes if the factor $s$ and $g$ at leat one is zero by using lemma (\ref{lemma}), i.e., the charge $\delta$ and cosmological constant $\Lambda$ are not allowed to arise together, which lead to the entropy sum is angular momentum independent under the consideration of theorem (\ref{theorem1}).

Finally, we arrive at the conclusion that the entropy sum of rotating charged(non-charged) Kaluza-Klein black holes and non-charged Kaluza-Klein-AdS black holes is angular momentum independent in even dimensions, while for odd dimensions, the angular momentum independence of the entropy sum holds for all general Kaluza-Klein black holes(no limits to the value of $\delta$ and $g$).

\subsection{Entropy Product}
The discussion of angular momentum independence of entropy product is a little more complex than that of entropy sum and our discussion is based on the lemma (\ref{lemma}) and theorem (\ref{theorem2}).

We introduce function
\begin{equation}
F(r)=\frac{(1+g^2r^2)^2}{4m\sqrt{f(r)}cr^{2-\epsilon}}\frac{r^{\epsilon-1}c}{\sqrt{f(r)}}\prod\limits_{j=1}^{n}\frac{r^2+a_j^2}{\chi_j}h(r)=
\frac{(1+g^2r^2)^2}{4mf(r)r^{3-2\epsilon}}\prod\limits_{j=1}^{n}\frac{r^2+a_j^2}{\chi_j}
\prod\limits_{k=1}^{2n+\lambda}(r-r_k)
\end{equation}
We take derivative of function $F(r)$ with respect to $r$, we get
\begin{equation}
\begin{split}
F'(r_i)&=\frac{(1+g^2r_i^2)^2}{4mf(r_i)r_i^{3-2\epsilon}}\prod\limits_{j=1}^{n}\frac{r_i^2+a_j^2}{\chi_j}
\prod\limits_{k\neq i}^{2n+\lambda}(r_i-r_k)\\
&=\frac{(1+g^2r^2)^2}{4m\sqrt{f(r)}cr^{2-\epsilon}}\frac{r^{\epsilon-1}c}{\sqrt{f(r)}}\prod\limits_{j=1}^{n}\frac{r^2+a_j^2}{\chi_j}\partial_r h(r)|_{r=r_i}\\
&=\frac{8\pi T_iS_i}{V_{d-2}}\label{9}
\end{split}
\end{equation}
together with E.q (\ref{5}) and (\ref{9}), we obtain
\begin{equation}
S_i T_i=\frac{V_{d-2}}{8\pi}\frac{(1+g^2r_i^2)}{2r_i^{1-\epsilon}\prod_{j=1}^{n}\chi_j}\prod\limits_{k\neq i}^{2n+\lambda}(r_i-r_k)\label{10}
\end{equation}
By using E.q (\ref{angular}), (\ref{5}) and (\ref{10}), we finally get
\begin{equation}
\sum_{i=1}^{2n+\lambda}\frac{\Omega_{ij}}{S_iT_i}=\frac{8\pi a_j\sqrt{\Xi_j}\prod_{j=1}^{n}\chi_j}{mcV_{d-2}}\sum_{i=1}^{2n+\lambda}
\frac{\prod_{k\neq j}^{n}(r_i^2+a_k^2)}{\prod_{k\neq i}^{2n+\lambda}(r_i-r_k)}\frac{1+g^2r_i^2}{f(r_i)r_i}\label{11}
\end{equation}

For E.q (\ref{11}), it's necessary to discuss it in four cases.
\begin{itemize}
\item Case \uppercase\expandafter{\romannumeral1}: $g=0$ and $\delta=0$\\
In this case, E.q becomes
\begin{equation}
\sum_{i=1}^{2n+\lambda}\frac{\Omega_{ij}}{S_iT_i}=\frac{8\pi a_j}{mV_{d-2}}\sum_{i=1}^{2n}
\frac{\prod_{k\neq j}^{n}(r_i^2+a_k^2)}{\prod_{k\neq i}^{2n}(r_i-r_k)}\frac{1}{r_i}\label{13}
\end{equation}
It's obviously to see that the maximal power term of factor $r_i^{-1}\prod_{k\neq j}^{n}(r_i^2+a_k^2)$ is $r_i^{2n-3}$ and the minimal power term is $\prod_{k\neq j}^{n}a_k^2r_i^{-1}$. By employing lemma (\ref{lemma}) and theorem (\ref{theorem2}), we can conclude that the entropy product of all horizons is angular momentum independent if there exist at least one rotation parameter $a_k=0$. Since the rotating black holes in four dimensions must own one rotation parameter, the conclusion we get can only be applied to black holes in dimensions $d\geq 5$, for charged(non-charged) rotating Kaluza-Klein black holes in four dimensions, the entropy product must be angular momentum dependent and it is straightforward to show that entropy product $\widehat S =a_1^2m^2/4$ in this case.
\end{itemize}

\begin{itemize}
\item Case \uppercase\expandafter{\romannumeral2}: $g=0$ and $\delta\neq0$\\
In this case, E.q (\ref{11}) becomes
\begin{equation}
\sum_{i=1}^{2n+\lambda}\frac{\Omega_{ij}}{S_iT_i}=\frac{8\pi a_j}{mcV_{d-2}}\sum_{i=1}^{2n}
\frac{\prod_{k\neq j}^{n}(r_i^2+a_k^2)}{\prod_{k\neq i}^{2n}(r_i-r_k)}\frac{1}{r_i}\label{12}
\end{equation}
The only difference between  E.q (\ref{12}) and (\ref{13}) is the extra constant parameter $c$ arises in (\ref{12}) which will not affect our discussion. So we have the same conclusion as that in case \uppercase\expandafter{\romannumeral1}
\end{itemize}

\begin{itemize}
\item Case \uppercase\expandafter{\romannumeral3}: $g\neq0$ and $\delta=0$\\
In this section, E.q (\ref{11}) becomes
\begin{equation}
\sum_{i=1}^{2n+\lambda}\frac{\Omega_{ij}}{S_iT_i}=\frac{8\pi a_j\prod_{j=1}^{n}\chi_j}{mV_{d-2}}\sum_{i=1}^{2n+2}
\frac{\prod_{k\neq j}^{n}(r_i^2+a_k^2)}{\prod_{k\neq i}^{2n+2}(r_i-r_k)}\frac{1+g^2r_i^2}{r_i}
\end{equation}
We could see that the maximal power term of factor $r_i^{-1}(1+g^2r_i^2)\prod_{k\neq j}^{n}(r_i^2+a_k^2)$ is $r_i^{2n-1}$ and the minimal power term is $\prod_{k\neq j}^{n}a_k^2r_i^{-1}$. By applying lemma (\ref{lemma}) and theorem (\ref{theorem2}), we can conclude that the entropy product is angular momentum independent if there exist at least one rotation parameter $a_k=0(d\geq5)$, just as the result in case \uppercase\expandafter{\romannumeral1} and \uppercase\expandafter{\romannumeral2}.
\end{itemize}

\begin{itemize}
\item Case \uppercase\expandafter{\romannumeral4}: $g\neq0$ and $\delta\neq0$\\
In this case, E.q (\ref{11}) has the form
\begin{equation}
\begin{split}
\sum_{i=1}^{2n+2}\frac{\Omega_{ij}}{S_iT_i}&=\frac{8\pi a_j\sqrt{\Xi_j}\prod_{j=1}^{n}\chi_j}{mcV_{d-2}}\sum_{i=1}^{2n+2}
\frac{\prod_{k\neq j}^{n}(r_i^2+a_k^2)}{\prod_{k\neq i}^{2n+2}(r_i-r_k)}\frac{1+g^2r_i^2}{f(r_i)r_i}\label{14}
\end{split}
\end{equation}
We carry out Tailor expansion for $1/f(r_i)r_i$
\begin{equation}
\frac{1}{f(r_i)r_i}=\frac{1}{(1-s^2g^2r_i^2)r_i}=\frac{1}{r_i}+g^2s^2r_i+g^4s^4r_i^3+...
\end{equation}
If we merely take consideration of lemma (\ref{lemma}), we could see that the value of E.q (\ref{14}) might be non-zero, even for the odd dimension(since the factor $(1+g^2s^2r_i^2)f(r_i)^{-1}r_i^{-1}\\\prod_{k\neq j}^{n}(r_i^2+a_k^2)$ is a function of $r_i$ with odd power, not even, which does not obey E.q (\ref{15})). But, the fact that the number/horizon radius $r_i$ are restricted by horizon function $h(r)$ which makes us unable to make sure whether the value of (\ref{14}) is zero or not, though it obviously does not meet the zero value condition stated in lemma (\ref{lemma}). In other words, lemma (\ref{lemma}) can only be applied to demonstrate the zero value of certain equations instead of non-zero value of the equations. Hence, the method we use fails in this case and it is unclear about the universal property of the entropy product/sum of the charged rotating Kaluza-Kleein-AdS black holes, though we believe more that the entropy product and entropy sum is angular momentum dependent than independent.
\end{itemize}

Through the argument above, we conclude that the entropy product of charged(non-charged) Kaluza-Klein black holes and non-charged Kaluza-Klein-AdS black holes is angular momentum independent if the black holes possess at least one zero rotation parameter $a_j=0$ in dimensions $d\geq 5$.

\section{Conclusions and Discussions}
Our study in this paper has generalized our discussions in ref.\cite{Hang} to general rotating Kaluza-Klein black holes, including the charged case, which is not discussed in ref.\cite{Hang}. We have investigated the angular momentum-independence of the entropy sum and entropy product of general rotating Kaluza-Klein black holes in higher dimensions by employing a mathematical lemma related to the Vandermonde determinant and the first law of thermodynamics for black holes. We have presented a general proof for the angular momentum-independence of the entropy sum and found the angular momentum free conditions for entropy product. We find that for charged(non-charged) rotating  Kaluza-Klein black holes and non-charged Kaluza-Klein-AdS black holes, the entropy sum is always angular momentum independent in dimensions $d\geq 4$, and the independence of angular momentum of the entropy product holds in higher dimensions $d\geq 5$ with at least one zero rotation parameter $a_j=0$. In  asymptotically flat spactime background, we have showed that the charge will not affect the angular momentum-independence of the entropy sum and product. For the non-charged case, the cosmological constant does not affect the angular momentum free property  of the entropy sum and product, too. Note that we have almost the same conclusions for the Myers-Perry-AdS black holes in all dimensions, as ref.\cite{Hang} has showed. While for the charged rotating Kaluza-Klein-AdS black holes, the universal property of the entropy sum and entropy product is unknown since the method we adopt fails due to the coupling effect of cosmological constant and charge and this problem is remained to be investigated in our future work.

According to the famous ``no hair or three hairs theorem for black hole", all the parameters we need to describe a black hole are just its mass $M$, charge $Q$ and angular momentum $J$. This simple property of black holes may give us a wonderful opportunity to study the thermodynamical property of the massive star by investigating its final simple state, i.e., black holes. For example, our study in this paper have showed that the entropy sum of the charged(non-charged) rotating Kaluza-Klein black holes and non-charged rotating Kaluza-Klein-AdS black holes is angular momentum free, this may give us a hint that the angular momentum is not so sensitive to the final entropy relations of the system, i.e., star/black holes. On the other hand, the existence of some  angular momentum  independent relations of a rotating star/black holes suggests that  these relations own universal property, which means no matter what initial conditions it starts the final state is universal.

Actually, in asymptotically (anti)-dS spacetime, the entropy sum of all horizons, including “virtual horizons”, only depends on the coupling constants of
the theory and topology of black holes, no matter in static spacetime or rotating spacetime for charged or uncharged black holes as shown in ref.\cite{Meng1,Meng2}, which suggests that the entropy sum is a ``real universal" relation and the mass, charge and angular momentum do not play an important role in entropy sum(though we have not generally proven this for all dimensions). For example, the entropy sum of the Kerr-Newman-AdS black hole is $S_1+S_2+S_3+S_4=-2l^2\pi$, where parameter $l$ is the cosmological radius.  The ``real universal" relations like entropy sum  are not only interesting but also profound, which may reflect the nature of black hole entropy and black hole thermodynamics. The other ``real universal" or universal relations can be found in ref.\cite{Hang,Xu2015,Meng2} and we will keep on looking for more the ``real universal" relations. We also hope that our study of the universal angular momentum-independence relations of the entropy sum and product, with the addition of mass-free relations in other cases, may help us a little bit to understand the the black hole entropy as it may imply new physics beyond the conventional considerations.

\section*{Acknowledgements}
For the present work we thank Xu Wei, Wang Deng and Yao Yanhong for helpful discussions. We would also like to thank the anonymous referee for helpful comments to make this work improved greatly. This project is partially supported by NSFC

\bibliography{Angular}
\end{document}